\documentclass[aps,twocolumn,prl,floatfix,showpacs,lengthcheck,hyperref,citeautoscript,superscriptaddress]{revtex4-1}
\usepackage{amsmath}
\usepackage{amssymb}
\usepackage{graphicx}
\usepackage{bm}
\usepackage{color}
\usepackage{times}
\usepackage{MnSymbol}
\usepackage{verbatim}

\newcommand{\bea}{\begin{eqnarray}}
\newcommand{\eea}{\end{eqnarray}}
\newcommand{\be}{\begin{equation}}
\newcommand{\ee}{\end{equation}}

\newcommand{\la}{\langle}
\newcommand{\ra}{\rangle}
\newcommand{\tav}{t_\text{av}}
\newcommand{\trel}{t_{\text{rel}}}
\newcommand{\ket}[1]{| #1 \rangle}

\newcommand{\bra}[1]{\langle #1 |}

\begin{document}
\title{Berry curvature tomography and realization of topological Haldane model in driven three-terminal Josephson junctions.}
\author{Lucila Peralta Gavensky}
\affiliation{Centro At{\'{o}}mico Bariloche and Instituto Balseiro,
Comisi\'on Nacional de Energ\'{\i}a At\'omica, 8400 Bariloche, Argentina}
\affiliation{Consejo Nacional de Investigaciones Cient\'{\i}ficas y T\'ecnicas (CONICET), Argentina}

\author{Gonzalo Usaj}
\affiliation{Centro At{\'{o}}mico Bariloche and Instituto Balseiro,
Comisi\'on Nacional de Energ\'{\i}a At\'omica, 8400 Bariloche, Argentina}
\affiliation{Consejo Nacional de Investigaciones Cient\'{\i}ficas y T\'ecnicas (CONICET), Argentina}

\author{D. Feinberg}
\affiliation{Centre National de la Recherche Scientifique, Institut NEEL, F-38042 Grenoble Cedex 9, France}
\affiliation{Université Grenoble-Alpes, Institut NEEL, F-38042 Grenoble Cedex 9, France}

\author{C. A. Balseiro}
\affiliation{Centro At{\'{o}}mico Bariloche and Instituto Balseiro,
Comisi\'on Nacional de Energ\'{\i}a At\'omica, 8400 Bariloche, Argentina}
\affiliation{Consejo Nacional de Investigaciones Cient\'{\i}ficas y T\'ecnicas (CONICET), Argentina}

\begin{abstract}
We propose a protocol to locally detect the Berry curvature of a three terminal Josephson junction with a quantum dot based on a synchronic detection when an AC modulation is applied in the device. This local gauge invariant quantity is expressed in terms of the instantaneous Green function of the Bogoliubov-de Gennes Hamiltonian. We analyze the contribution to the Berry curvature from both the quasi-particle excitations and the Andreev bound state levels by introducing an effective low-energy model. In addition, we propose to induce topological properties in the junction by breaking time-reversal symmetry with a microwave field in the non-resonant regime. In the last case, the Floquet-Andreev levels are the ones that determine the topological structure of the junction, which is formally equivalent to a 2D-honeycomb Haldane lattice. A relation between the Floquet Berry curvature and the transconductance of the driven system is derived.
\end{abstract}
\maketitle
\textit{Introduction}.-- 
Multiply connected electronic networks threaded by flux tubes have been proposed several years ago as a platform to develop quantized adiabatic transport properties 
intimately related to topological invariants~[\onlinecite{Avron1988}] with possible realizations in Josephson junctions~[\onlinecite{Avron1989}]. More recently, band structures of Andreev bound states (ABS) in $N$-terminal Josephson junctions of conventional superconductors have been shown to host topological singularities for $N\geq 4$, such as zero-energy Weyl points, in the artificial reciprocal lattice space defined by the $N-1$-independent superconducting phases~[\onlinecite{Riwar2016},\onlinecite{Eriksson2017}]. Even more, when adding a magnetic flux through the central region and hence breaking time reversal symmetry, tri-junctions may also realize non-trivial topology~[\onlinecite{Meyer2017},\onlinecite{Xie2017}]. The topological structure of these devices can be probed by means of transconductance measurements between two voltage biased terminals which yields, at vanishing voltage, a quantized value proportional to the first Chern number of the ground state~[\onlinecite{Avron1989},\onlinecite{Riwar2016},\onlinecite{Eriksson2017}]. This global topological invariant involves an integral over phase space of a gauge invariant geometric magnitude, the Berry curvature. A non-zero Berry curvature manifests itself in physical effects, such as anomalous velocities, regardless if whether the Chern number is non-trivial~[\onlinecite{Sundaram1999},\onlinecite{Xiao2010}]. In fact, providing experimental tools to reconstruct maps of the curvature has been a noteworthy work over the past few years~[\onlinecite{Flaschner2016,Li2016,Wimmer2017}].    

Manifestations of topology in Josephson junctions of conventional superconductors are still an open case of study. An interesting question to address is whether it is possible to induce topological properties on a three terminal device by introducing a periodic driving in the superconducting phases. Indeed, non-trivial Berry curvatures can take place within the picture provided by the Floquet states. The purpose of this work is then two-fold. First, we propose an experimentally suitable protocol to locally measure the Berry curvature of the ground state wavefunction of the junction. In order to do so, we exploit one of the main advantages of these mesoscopic setups: the potential to control the superconductor phases so as to perform local transport measurements at each point of the artificial Brillouin zone. Secondly, we study the topology of Floquet-Andreev bands in the large frequency limit and provide a relation between their curvature and the transconductance of a biased and periodically driven setup. This constitutes a novel  realization of the Haldane model in a solid state device~[\onlinecite{Haldane1988}], becoming an alternative to previous achievements in cold atom systems~[\onlinecite{Jotzu2014}]. 
\begin{figure}[t]
\includegraphics[width=0.8\columnwidth]{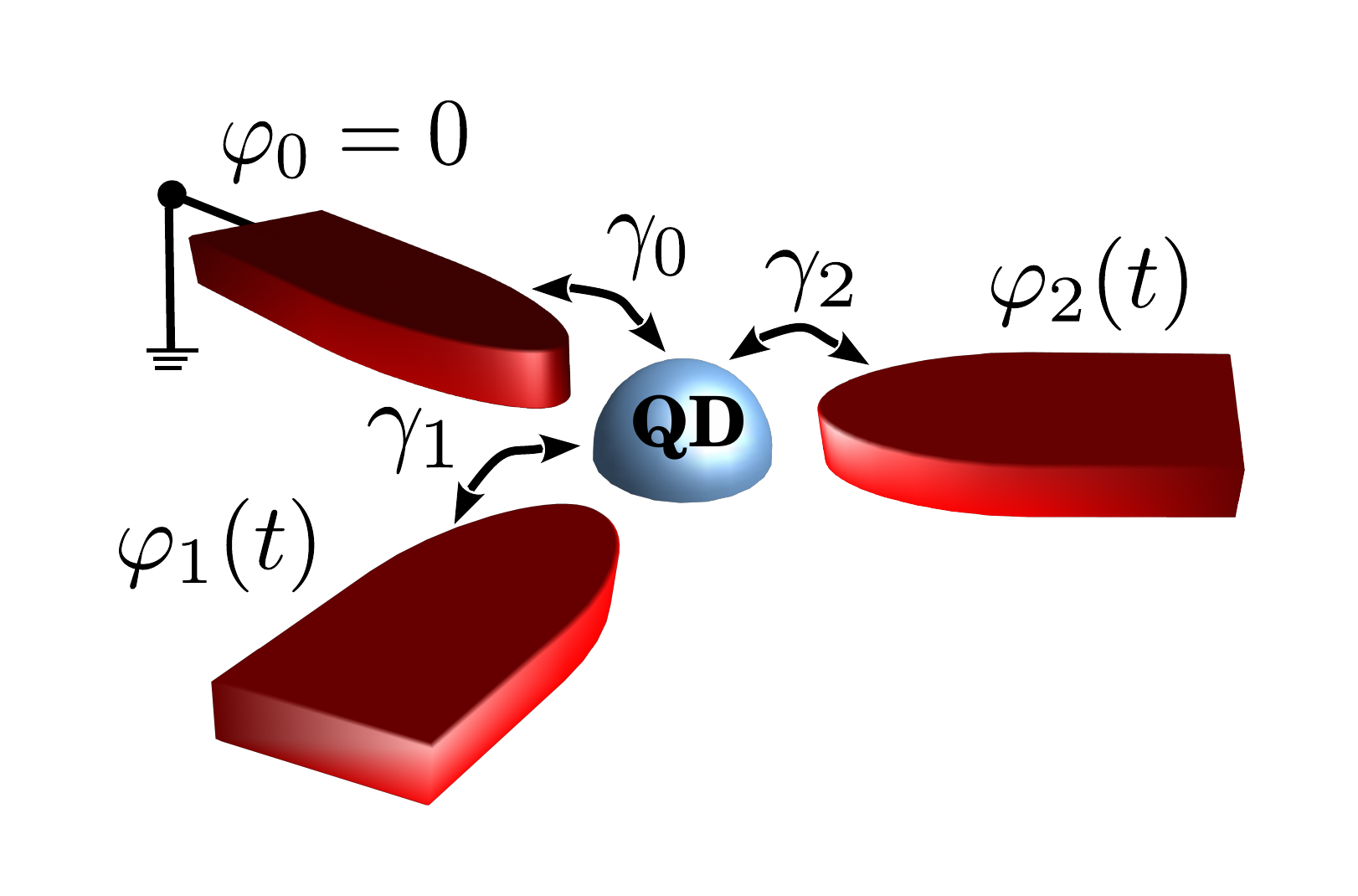}
\caption{General scheme of a three terminal Josephson junction with a quantum dot. Two of the superconducting phases are changed in time by voltage biasing the leads or varying the external flux in a ring setup.}
\label{fig1}
\end{figure}

\textit{Model Hamiltonian and synchronic measurement of the Berry curvature.}--
We begin by studying a three-terminal Josephson junction (3TJJ) with a quantum dot bearing a single relevant level in the energy gap region of the superconductor (see Fig. \ref{fig1}). Its Hamiltonian, neglecting Coulomb interactions, can be written as
\begin{eqnarray}
\label{eq1_p}
\mathcal{H}(t) &=& \sum_{\bm{k}\sigma\nu}\xi^{}_{\bm{k}\nu}c^{\dagger}_{\bm{k}\sigma\nu}c^{}_{\bm{k}\sigma\nu} - \Delta\sum_{\bm{k}\nu}(c^{\dagger}_{\bm{k}\uparrow\nu}c^{\dagger}_{\bm{-k}\downarrow\nu} + h.c)\\
\notag
&&+\varepsilon^{}_d\sum_{\sigma}d^{\dagger}_{\sigma}d^{}_{\sigma}+\sum_{\bm{k}\sigma\nu}\Big(\gamma_{\nu}e^{-i\varphi_{\nu}(t)/2}d^{\dagger}_{\sigma}c_{\bm{k}\nu\sigma}^{} + h.c.\Big),
\end{eqnarray}
where we defined $\xi_{\bm{k}\nu} = \varepsilon_{\bm{k}\nu} - \mu$ and the superconducting phases have been gauged to the tunneling matrix e\-le\-ments between the leads and the dot. Due to gauge invariance, one of the phases is chosen to be zero without loss of generality.
Even though the band structure of the ABS in equilibrium remains topologically trivial, its  Berry curvature, defined in the $(\varphi_1, \varphi_2)$ plane is non-zero provided $\varepsilon_d \neq 0$ due to breaking of particle-hole symmetry. Our first approach is to study the transport properties of the junction when making adiabatic variations of the fluxes threading the nanostructure. To this end, we use Keldysh formalism [\onlinecite{Jauho1996},\onlinecite{Cuevas1996}] working in Nambu space, where the spinors in the leads and the dot are defined as $\Psi^{\dagger}_{\bm{k}\nu}(t) = \left(c^{\dagger}_{\bm{k}\nu\uparrow}(t), c^{}_{-\bm{k}\nu\downarrow}(t)\right)$ and $\Psi^{\dagger}_{d}(t) = \left(d^{\dagger}_{\uparrow}(t), d^{}_{\downarrow}(t)\right)$, respectively. The current flowing into the reservoir lead $\nu$ can be expressed as
\begin{eqnarray}
\langle\phi_{0}|\mathcal{J}_{\nu}(t)|\phi_{0}\rangle =\frac{2e}{\hbar}\Re\text{Tr}\Big[\sigma_z\hat{V}_{\nu d}(t)\hat{\mathcal{G}}^{<}_{d\nu}(t,t)\Big]\,,
\label{eq2_p}
\end{eqnarray}
where the trace is performed in Nambu space, $[\mathcal{G}_{d\nu}^{<}]^{\alpha\beta}(t,t') = i\langle\phi_0|\Psi^{\dagger\alpha}_{d}(t')\Psi^{\beta}_{\nu}(t)|\phi_0\rangle$ are the components of the minor Green function at the link between the dot and lead $\nu$ and the time dependent tunneling is given by $\hat{V}_{\nu d} = \gamma_{\nu}e^{i\frac{\varphi_{\nu}(t)}{2}}(\mathcal{I}+\sigma_{z})/2 - \gamma_{\nu}e^{-i\frac{\varphi_{\nu}(t)}{2}}(\mathcal{I}-\sigma_{z})/2$. In order to perform an adiabatic expansion we will work in the Wigner representation of the two-time Green functions: $\tilde{\mathcal{G}}(\omega,\tav) = \int_{-\infty}^{\infty} d\trel e^{i\omega\trel}\hat{\mathcal{G}}(t,t')$, where we introduced  the relative time $\trel = t-t'$ and the average one $\tav = (t+t')/2$. Time
derivatives of the latter are used as a small parameter, making feasible a perturbation scheme. The corrections up to first order of the current expectation value are found to be~\cite{note1}
\begin{eqnarray}
\notag
\langle\!\mathcal{J}_{\nu}(t)\!\rangle\!&=&\!2e\lim_{t'-t\to\epsilon^{+}}\!\left\{\!-\frac{i}{\hbar}\!\int\!\frac{d\omega}{2\pi}\!\mathrm{Tr}\left[\!\frac{\partial H}{\partial\varphi_{\nu}}\tilde{\mathcal{G}}^{c}_{0}\!\right]\!\right.\\
\nonumber
\!&-&\!\left.\!\sum_{\rho}\!\!\int\!\!\frac{d\omega}{4\pi}\!\mathrm{Tr}\left[\!\epsilon^{\nu\rho}  {\tilde{\mathcal{G}}^{c^{-1}}_{0}}\!\frac{\partial {\tilde{\mathcal{G}}^{c^{}}_{0}}}{\partial{\varphi_{\nu}}}\!\cdot {\tilde{\mathcal{G}}^{c^{-1}}_{0}}\!\frac{\partial {\tilde{\mathcal{G}}^{c^{}}_{0}}}{\partial{\varphi_{\rho}}}\!\cdot {\tilde{\mathcal{G}}^{c^{-1}}_{0}}\!\frac{\partial {\tilde{\mathcal{G}}^{c^{}}_{0}}}{\partial{\omega}}\!\right]\!\dot{\varphi}_{\rho}\!\right\}\\
\label{eq3_p}
\!&=&\frac{2e}{\hbar}\frac{\partial\varepsilon_g(t)}{\partial\varphi_{\nu}}- 2e\sum_{\rho}\mathcal{F}^{g}_{\nu\rho}(t)\dot{\varphi}_{\rho}(t),
\end{eqnarray}
where $ {\tilde{\mathcal{G}}^{c^{}}_{0}}(\omega,\tav) = [\omega - H(\tav)]^{-1}$ is to be understood as the causal Green function of the Hamiltonian defined in Eq. (\ref{eq1_p}) in the total adiabatic limit, that is to say, the pro\-pa\-gator that follows the perturbation  instantaneously. In the last equality we made the identification of the zeroth order mean value of the current with the phase derivative of the instantaneous ground state energy, along with the representation of the Berry curvature of the ground state $\mathcal{F}^{g}_{\nu\rho}$ in terms of the single-particle time ordered Green's function of the problem~[\onlinecite{Niu1985,Gurarie2011,Wang2012}].
We therefore recover the results discussed in Ref. [\onlinecite{Riwar2016}] with the virtue of identifying a way of obtaining adiabatic corrections, taking into account the continuum states in full extent, without the need of working with the ground state-wave function. 

The synchronic detection protocol to measure the Berry curvature is based on performing a periodic modulation in one of the fluxes that fixes the superconductors phase difference while keeping the other one constant. We take, without loss of generality, $\varphi_{1} = \varphi_{1}^{0}$ and $\varphi_2(t) =\varphi_{2}^{0}+ b\sin(\mathcal{V} t)$. Assuming the amplitude of the driving field to be small $b\ll1$ and the adiabatic postulate to hold, the current in lead $\nu=1$ takes the form
\begin{eqnarray}
\notag
\la\!\mathcal{J}_{1}(t)\!\ra\!&\simeq&\!\frac{2 e}{\hbar}\!\left(\!\frac{\partial\varepsilon_{g}}{\partial \varphi_{1}}\!+\!\frac{\partial^2\varepsilon_{g}}{\partial \varphi_{2}\partial\varphi_{1}}b\sin(\mathcal{V}t)\!\right)\biggr|_{\bm{\varphi}^{0}}\\
&\!-\!&2e\mathcal{F}_{12}^{g}\biggr|_{\bm{\varphi}^{0}}b\, \mathcal{V}\cos(\mathcal{V} t)\,,
\end{eqnarray}
where we neglected terms of $\mathcal{O}(b^2)$.
Hence, the zeroth and first order corrections are in quadrature with each other. A synchronic filter of the current flowing into the reservoir lead with the derivative of the signal can be performed, leading to a local measurement of the Berry curvature, since 
\begin{equation}
\frac{\mathcal{V}}{2\pi}\int_{0}^{\frac{2\pi}{\mathcal{V}}}\langle\hat{J}_{1}(t)\rangle\cos(\mathcal{V} t)dt = -e\mathcal{F}_{12}^{g}|_{\bm{\varphi}^{0}}b\mathcal{V}\,.
\label{eq5_p}
\end{equation}
In order to compare the full numerical results with a simpler model, we introduce a low energy approximation of the ABS~[\onlinecite{Bauer2007,Affleck2000,Vecino2003}], with an effective Hamiltonian given by
\begin{equation}
\hat{H}_{\text{eff}}(\bm{\varphi})\!=\!-\hat{G}^{-1}_{dd}(\omega \!=\! 0)\!=\!\left(\begin{array}{cc}
\varepsilon_{d}\! &\!\sum_{\nu}\!\Gamma_{\nu} e^{-i\varphi_{\nu}}\\
\sum_{\nu}\!\Gamma_{\nu}e^{i\varphi_{\nu}}&-\varepsilon_{d}
\end{array}\right),
\label{eq6_p}
\end{equation}
where the anomalous on-site interaction is expressed in terms of the hybridization $\Gamma_{\nu} = -{\gamma}^2_{\nu}\rho(\varepsilon_F)\pi$. This limit is expected to be generally accurate whenever $\gamma_{\nu}\ll\Delta$, so that the bound states possess less hybridization with the continuum above the gap. It is then possible to define an effective Berry curvature given by
\begin{equation}
\mathcal{F}_{21}^{\text{eff}}(\bm{\varphi}) = \frac{1}{2\pi} \bm{\hat{h}}_{\text{eff}}\cdot(\partial_{\varphi_2}\bm{\hat{h}}_{\text{eff}}\times\partial_{\varphi_1}\bm{\hat{h}}_{\text{eff}}),
\label{eq7_p}
\end{equation}
with $\bm{h}_{\text{eff}} = (\Gamma_0 + \Gamma_1\cos(\varphi_1) + \Gamma_2\cos(\varphi_2), \Gamma_1\sin(\varphi_1) + \Gamma_2\sin(\varphi_2), \varepsilon_d)$ and $\bm{\hat{h}}_{\text{eff}} = \bm{h}_{\text{eff}}/|\bm{h}_{\text{eff}}|$.
\begin{figure}[t]
\includegraphics[width=\columnwidth]{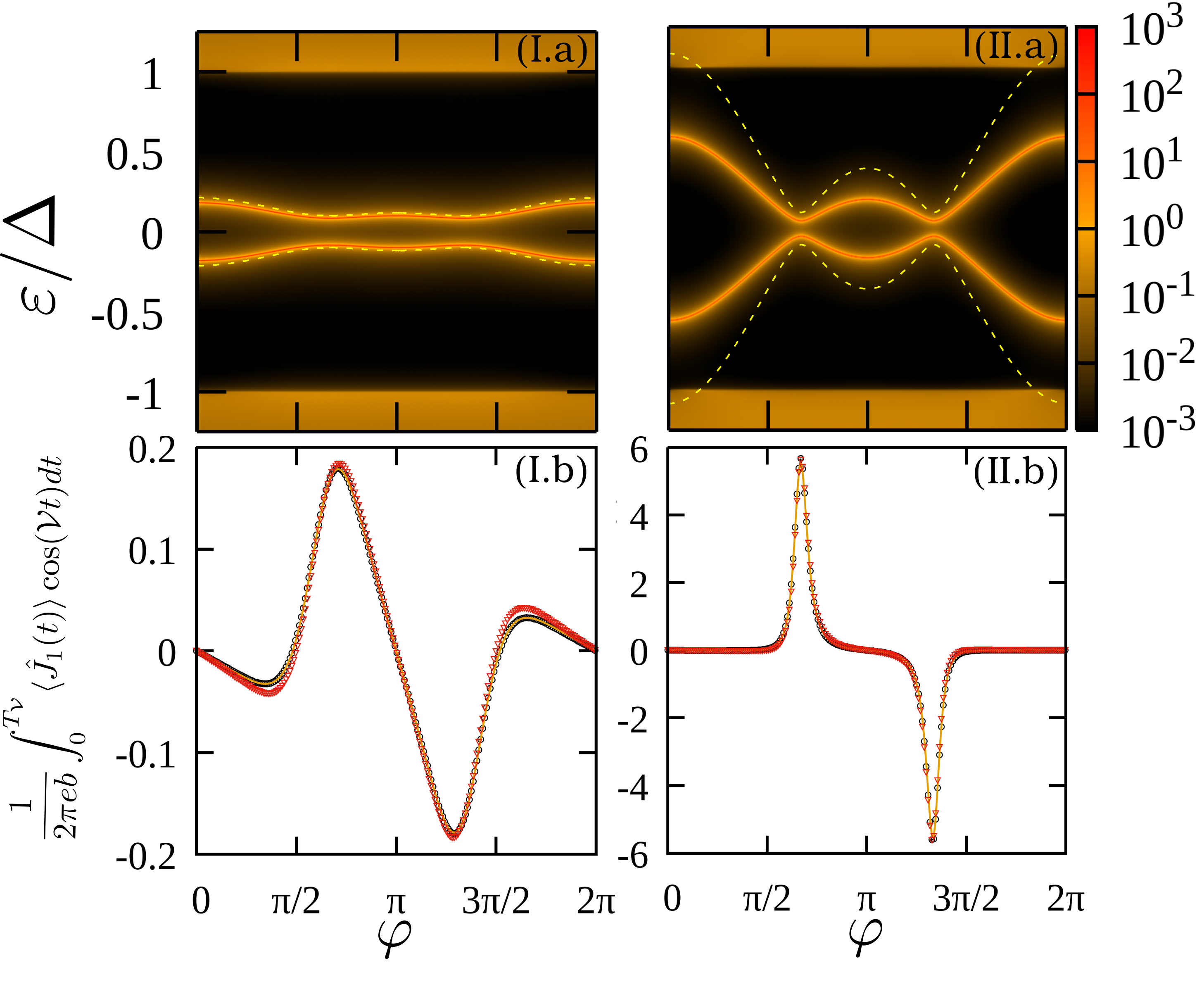}
\caption{Upper panels show the dot spectral density in units of $\Delta$ along $\varphi=\varphi_1 = -\varphi_2$ for a symmetric 3TJJ in equilibrium with a hopping amplitude to the leads $\gamma=0.25\Delta$ \textbf{(I.a)} and $\gamma=0.6\Delta$ \textbf{(II.a)}. The dot energy was taken to be $\varepsilon_d = 0.1\Delta$ in both cases. Dashed lines indicate the dispersion of Andreev levels captured by the infinite gap limit. Lower panels show the outcome of the filter of the $\cos(\mathcal{V}t)$ component of the current flowing into the bia\-sed lead $\nu = 1$ obtained using Keldysh formalism (open circles) and its comparison with both the Berry curvature of the exact Hamiltonian $\mathcal{F}_{21}^{g}$ [Eq. (\ref{eq3_p})] (solid line) and the one obtained with
Eq. (\ref{eq7_p}) (triangles).}
\label{fig2}
\end{figure}
Interestingly, the Hamiltonian in Eq.~(\ref{eq6_p}) enjoys all the topological information of a honeycomb lattice in a tight-binding description, identifying $\varphi_{\nu} = \bm{k}\cdot\bm{a}_{\nu}$ with $\bm{a}_{\nu}$ the displacements of a site respect to its three nearest neighbors and $2\varepsilon_d$ the subblatice energy difference. In what follows we perform an exact calculation of the current as expressed in Eq.~(\ref{eq2_p}) taking into account the perturbation to all orders by making use of the time periodicity of the Green functions~\citep{note1} and compare its synchronic filter with the Berry curvature.
In Fig. \ref{fig2} results for a symmetric 3TJJ ($\gamma_{\nu} = \gamma$) are shown. The dot spectral density $A_{d}(\omega) = -(1/\pi)\Im\text{Tr}(G^r_{dd})$ is depicted for different values of $\gamma/\Delta$ along the high-symmetry path $\varphi_1 = -\varphi_2$, finding a good agreement with the bands of $\hat{H}_{\text{eff}}$ (indicated with dashed lines) for $\gamma=0.25\Delta$ [Fig. \ref{fig2}\,\textbf{(I.a)}]. As expected, this description turns out to be insufficent for $\gamma=0.6\Delta$ [Fig. \ref{fig2}\,\textbf{(II.a)}]. In both cases the dot energy is $\varepsilon_d = 0.1\Delta$. Lower pannels show the filter of the current flowing into the reservoir lead $\nu=1$ as obtained from Eq. (\ref{eq5_p}) and its comparison with both $\mathcal{F}_{21}^{g}$ [see Eq. (\ref{eq3_p})] and the low energy approximation of this quantity given by the effective model $\mathcal{F}^{\text{eff}}_{21}$ [Eq. (\ref{eq7_p})]. In the spirit of the adiabatic approximation to be valid, the frequency was chosen to be $\mathcal{V} = 10^{-2}\Delta$ and the field amplitude $b=10^{-4}\Delta$. We observe that the synchronic protocol gives a precise information on the curvature of the Andreev bands captured by the low energy theory even when the hybridization with the states above the superconductor gap is relevant. We can comprehend this behavior by noticing that the Berry curvature in this model is a low-energy localized quantity. In both cases, even though the integral over the whole phase space of the Berry curvature is zero, the current has a topological Thouless contribution whenever the phases of the junction are near the Dirac points.

\begin{figure}[t]
\includegraphics[width=\columnwidth]{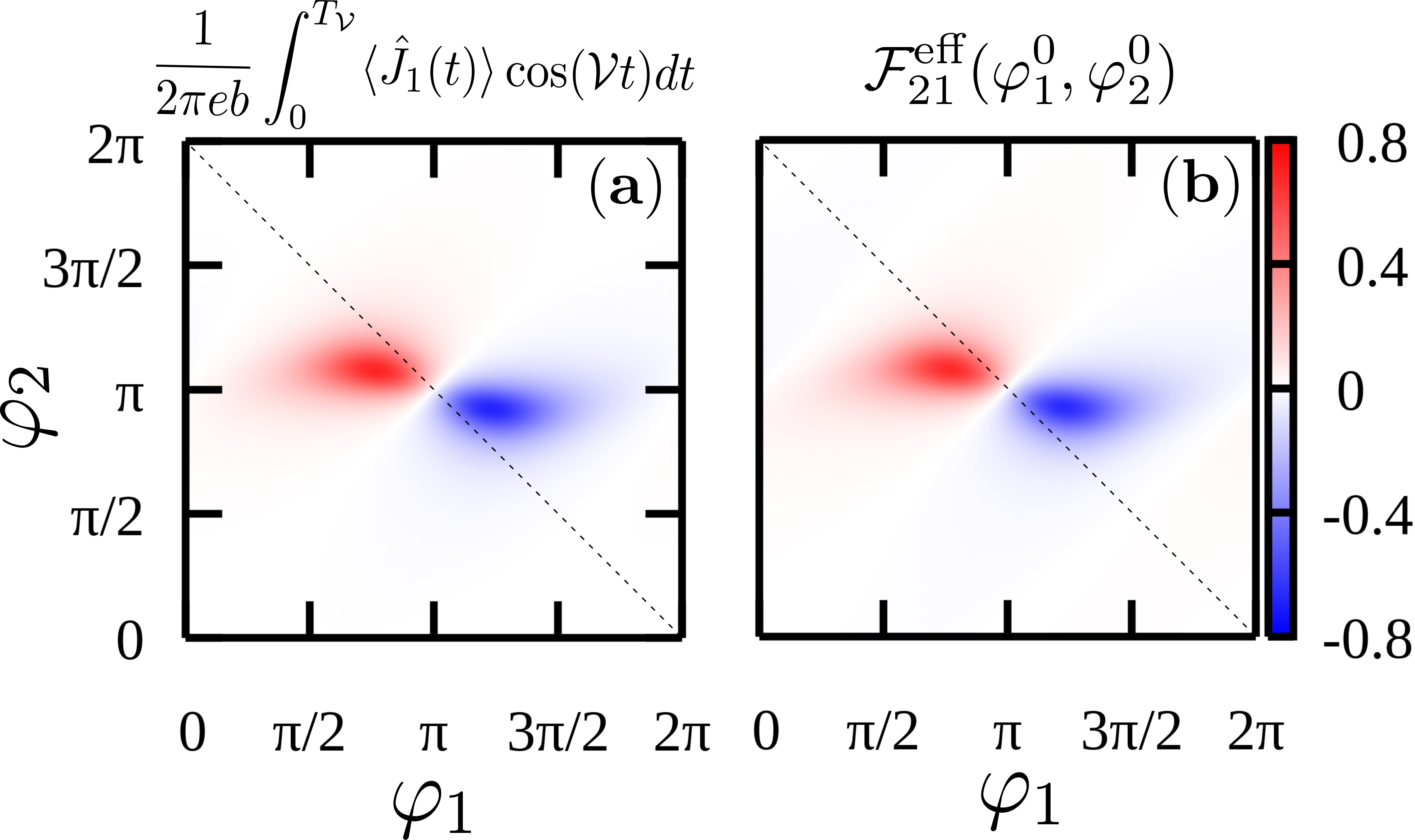}
\caption{\textbf{(a)} Synchronic filter of the of the current flowing into the biased lead $\nu=1$ as obtained from Eq.\,(\ref{eq5_p}) using Keldysh formalism throughout the whole phase space. \textbf{(b)} Berry curvature obtained within the low energy effective Hamiltonian describing the infinite gap limit.}
\label{fig3}
\end{figure}

This procedure can be done along the whole phase space, so as to obtain a map of the Berry curvature with the proposed transport measurement. In Fig. \ref{fig3} we show the comparison for an asymmetric junction with hoppings $\gamma_1=0.3\Delta$, $\gamma_2 = 0.5\Delta$ and $\gamma_3 = 0.6\Delta$ of the low energy model with the results of the finite-gap Hamiltonian using Keldysh formalism. We find an excellent agreement between both, that guarantees the validity of the approximation given by Eq. (\ref{eq6_p}). We thus conclude that the effect of the continuum states on the topological properties of this model is negligible.
 It is worth emphasizing however, that our expression for the transconductance (second term in  Eq.~(\ref{eq3_p}))  reproduces the correct quantized average value  \cite{lupe2018} even when the continuum does contribute as in the model presented in Ref.~\cite{Meyer2017}.

\textit{Inducing topology by driving: Floquet-Andreev physics.}-- When introducing a periodic driving that breaks time reversal symmetry (TRS) into the three terminal junction, the resulting Floquet system can have topological properties which differ from the original one. This can be engineered by introducing an ``elliptically" polarized driving~[\onlinecite{Venitucci2017}] of the form $\delta\varphi_{\nu}(t) = A_0\cos(\Omega t + \chi_{\nu})$ with $\chi_1 - \chi_2$ not a $\pi$-multiple. The Floquet Hamiltonian of the system is given by
$
[\mathcal{H}_F]_{mn} = (H_0 - m\Omega)\delta_{mn} + \mathcal{U}_{m-n},
$
with $H_0$ representing the undriven Hamiltonians of the dot and leads and $\mathcal{U}_{m-n} = 1/T\int_{0}^{T}e^{i(m-n)\Omega t} \mathcal{U}(t)$ the $m-n$'th harmonic of the time-dependent tunneling. The Floquet Green's functions satisfy $[\omega - \mathcal{H}_{F}]{\mathcal{G}}^{F}(\omega) = \mathcal{I}$, and its matrix elements can be used to reconstruct the full two-time dependent original Green's function
\begin{equation}
\check{\mathcal{G}}(t,t') = \sum_{m,n}\int_{0}^{\Omega}\frac{d\omega}{2\pi}e^{-i(\omega + m\Omega)t}e^{i(\omega + n\Omega)t'}\mathcal{G}^{F}_{mn}(\omega)
\end{equation}   
\begin{figure}[t]
\includegraphics[width=\columnwidth]{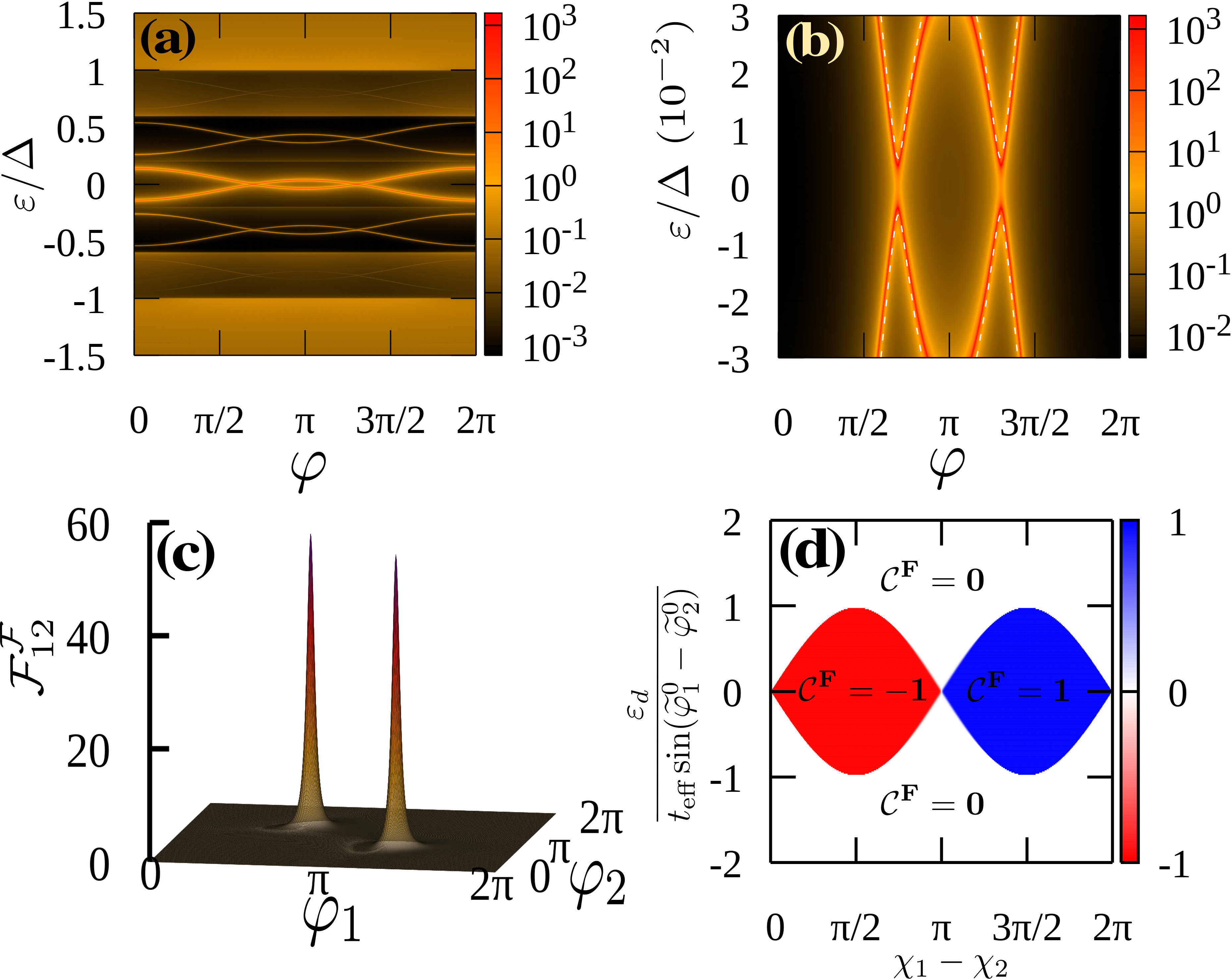}
\caption{\textbf{(a)} Spectral density of the Floquet Green function of the dot projected onto the zeroth replica for $\Omega = 0.4\Delta$, $A_0=0.8$ and $\varepsilon_d = 0$. \textbf{(b)} Zoom of the spectral density near zero energy, a gap opens due to breaking of TRS. Dashed lines indicate the dispersion relation of the Floquet eigenergies obtained within the \textit{infinite gap} approximation. \textbf{(c)} Berry curvature of the Floquet-Andreev band in phase space with parameters as in \textbf{(a)}. \textbf{(d)} Floquet Chern numbers of the junction as a function of the polarization of the driving.}
\label{fig4}
\end{figure}

In Fig.~\ref{fig4}\textbf{(a)} we show the spectral density of the Floquet
Green's function of the dot projected onto the zeroth replica $A_d^{F} = -(1/\pi)\Im\text{Tr}_{00}[{\mathcal{G}}^{F}_{dd}(\omega)]$ for $\chi_1 - \chi_2 = \pi/2$, $\varepsilon_d = 0$, $\Omega = 0.4\Delta$ and $A_0=0.8$. In this case symmetric hoppings $\gamma_{\nu} = 0.25\Delta$ were
chosen and a representation along the path $\varphi_2 = -\varphi_1$ is shown. The breaking of TRS removes the degeneration at the diabolical Dirac points, producing the appearance of a zero energy gap in the Floquet spectrum. In Fig.~\ref{fig3}\textbf{(b)} we show a zoom of the spectral density near the Fermi level along with a comparison with the Floquet eigenergies obtained from the effective low-energy theory, expected to remain suitable for these parameters. Indeed, since the energy of the microwave field is larger than the bandwith of the Andreev bands the only relevant topological change can take place at zero energy, since no  gaps  at the first Floquet zone boundary ($\varepsilon = \pm\hbar\Omega/2$) can be present. On the other hand, the first Floquet replicas of the Andreev band structure lie within the superconductors gap so that the continuum states will renormalize only perturbatively  the dispersion relation of the Floquet bands at low energy. 
In Fig.~\ref{fig4}\textbf{(c)} the Berry curvature of the lower Floquet-Andreev band obtained with the driven e\-ffec\-tive model is shown. In this case both Weyl points contribute with the same topological charge leading to a non-trivial Chern number. Standard Brillouin-Wigner perturbation theory yields, in this regime, a Haldane-like effective Floquet Hamiltonian on the projected zero photon subspace~[\onlinecite{Kitagawa2011},\onlinecite{Mikami2016}], with renormalized parameters of the form 
\begin{eqnarray}
\widetilde{\varepsilon}_d(\bm{\varphi^0}) &=& \varepsilon_d - t_{\text{eff}}\sin(\varphi_1^0 - \varphi_2^0)\sin(\chi_1 - \chi_2)\,,\\
\notag
\widetilde{\Gamma}_{1(2)} &=& \Gamma_{1(2)}J_0(A_0),
\end{eqnarray} 
and $t_{\text{eff}} = 4\Gamma_1\Gamma_2J_1^2(A_0)/\hbar\Omega$ acting as a second neighbor effective hopping in the previously mentioned honeycomb lattice description. The Dirac points in the symmetric model ($\Gamma_{\nu} = \Gamma$) are displaced from their original position along the path $\widetilde{\varphi}_2^{0} = -\widetilde{\varphi}_1^{0}$ such that $\cos(\widetilde{\varphi}_1^{0}) = -1/2J_0(A_0)$. The topological phase diagram of the driven tri-junction is shown in Fig.~\ref{fig4}\textbf{(d)} as a function of polarization, which measures the breaking of TRS. The dot energy has been normalized to the value it takes when the gap closes for ``circularly" polarized driving.  

A crucial issue concerns the possibility of following adiabatically the Floquet states in order to accurately probe their properties. The equilibrium populations are strongly modified whenever a gap opens in the spectrum so that the most probable scenario is that the wavefunction at these singular phases behaves as a superposition of Floquet states, leading to a non-thermal occupation of the Floquet bands. Nonetheless, if the phases of the superconductors are setup far away from the latter there is a precise adiabatic connection between the unperturbed Hamiltonian and the dressed Floquet-Andreev bands. 
We now show that voltage biasing these adiabatically connected states can provide information on the Berry curvature of the Floquet bands. In order to do so we appeal to the two-time formalism~[\onlinecite{Peskin1993,Pfeifer1983,Drese1999}]. Under this formulation, an extended Schr$\ddot{\text{o}}$dinger equation is taken into account, 
$
[\mathcal{H}(t,\tau) - i\hbar\partial_t]|\psi(t,\tau)\ra\ra = i\hbar\partial_{\tau}|\psi(t,\tau)\ra\ra,
$
where the time variable $t$ is associated with the fast time-periodic evolution of frequency $\Omega$, while $\tau$ accounts for the slow time dynamics generated by the bias voltages at the leads, i.e. the phases are such that $\varphi_{\nu}(t,\tau) = \theta_{\nu}(t) + \phi_{\nu}(\tau)$, with $\theta_{\nu}(t) = A_0\cos(\Omega t + \lambda_\nu)$ and $\phi_{\nu}(\tau) = \varphi_{\nu}^{0} + \frac{2e}{\hbar}V_{\nu}\tau $. The notation $|...\rangle\rangle$ indicates that the two-time wavefunctions are to be treated in a Hilbert space with the internal product $\langle\langle \psi^{\alpha}(\tau)|\psi^{\beta}(\tau)\rangle\rangle = \frac{1}{T}\int_{0}^{T}\langle \psi^{\alpha}(\tau,t)|\psi^{\beta}(\tau,t)\rangle dt$. The adiabatic basis of the problem is the Floquet basis, meaning that these are the states $|u_{\beta}^{F}(t,\tau)\ra\ra$ that satisfy the instantaneous Floquet equation 
$
[\mathcal{H}(t,\tau)-i\hbar\partial_t]|u^F_{\beta}(t,\tau)\ra\ra = \varepsilon^F_{\beta}(\tau)|u^F_{\beta}(t,\tau)\ra\ra,
$
where $\beta = 1,2$ indexes the two bands of the first Floquet zone with quasi-energies $-\hbar\Omega/2 < \varepsilon_{\beta}^{F} < \hbar\Omega/2$. When treating the bias as an adiabatic perturbation in the slow scale, the mean value of the current operator at each lead is found to be
\begin{equation}
\la\la\mathcal{J}_{\nu}\ra\ra_{\alpha} = \frac{2 e}{\hbar} \frac{\partial\varepsilon^F_{\alpha}}{\partial \varphi_{\nu}}[\bm{\phi}(\tau)]-\frac{4e^2}{\hbar}\sum_{\rho}\mathcal{F}_{\nu\rho}^{F\alpha}[\bm{\phi}(\tau)]V_{\rho},
\label{eq9_p}
\end{equation} 
where the first term corresponds to the adiabatic Floquet-Josephson supercurrent and the second accounts for the Berry curvature of the $\alpha$-Floquet state 
\begin{equation}
\mathcal{F}_{\nu\rho}^{F\alpha} = i[\la\la\partial_{\varphi_{\nu}}u^{F}_{\alpha}|\partial_{\varphi_{\rho}}u^{F}_{\alpha}\ra\ra-\la\la\partial_{\varphi_{\rho}}u^{F}_{\alpha}|\partial_{\varphi_{\nu}}u^{F}_{\alpha}\ra\ra].
\end{equation}
We arrive to an expression for the currents with a transconductance term proportional to the gauge field, in analogy with the results obtained for undriven setups in Ref.~[\onlinecite{Riwar2016}]. If incommensurate voltages were to be applied, so that the entire phase space is probed after a considerable time, then Eq. (\ref{eq9_p}) results in a topologically quantized transconductance since 
\begin{equation}
\int_{BZ}\frac{d^2\varphi}{(2\pi)^2}\la\la\mathcal{J}_{\nu}\ra\ra_{\alpha} = -\frac{4e^2}{h}\sum_{\rho}C^{F\alpha}_{\nu\rho}V_{\rho},
\end{equation}
where $C^{F\alpha}_{\nu\rho}$ is the Chern number of the Floquet band $\alpha$. 

\textit{Conclusion.}-- We have proposed a protocol to perform local measurements of the Berry curvature of the ground-state wavefunction of Josephson junctions. This gauge invariant quantity [Eq.~(\ref{eq3_p})] is obtained with the Feynman propagator of the Hamiltonian which includes the quasi-particle excitations and compared with a low energy effective model of the ABS spectrum. We found a good agreement between both since the gauge field of the model is localized at energies close to zero. On the other hand, we have put forward a way of inducing topological properties on three terminal devices by introducing a periodic driving that breaks time reversal symmetry and generates dressed Floquet-Andreev bands with non-trivial Haldane-like Chern numbers. We have proposed that the transconductance of the driven junction yields information of the Floquet Berry curvature when voltage biasing states adiabatically connected to the Floquet bands. 

We acknowledge financial support from PICTs 2013-1045 and 2016-0791 from ANPCyT, PIP 11220150100506 from CONICET and grant 06/C526 from SeCyT-UNC. 

\begin{widetext}
\section{SUPPLEMENTARY MATERIAL}

\subsection{1.Floquet-Dyson equations for green functions of periodically driven Hamiltonians.}
The Green's functions of a $T$-periodically-driven system depend on two independent time variables, even in the steady state. The Dyson equation for the Green's functions denoted in Keldysh representation is given by
\begin{equation}
\check{\mathcal{G}}(t,t') = \check{g}(t-t') + \int \check{g}(t-t'')V(t'')\check{\mathcal{G}}(t'',t')dt''.
\label{eq1}
\end{equation}
In order to take advantage of the discrete time symmetry of the Hamiltonian, we introduce the Wigner-transformation, defining the relative time $t_{\text{rel}} = t-t'$ and the average $t_{\text{av}} = \frac{t+t'}{2}$. For an arbitrary Green's function satisfying the periodicity condition on each of its time arguments $\check{\mathcal{G}}(t,t') = \check{\mathcal{G}}(t+T,t'+T)$, its invariance on $t_{\text{av}}\rightarrow t_{\text{av}} + T$ is also guaranteed. We can then write
\begin{equation}
\check{\mathcal{G}}(t,t')\!=\!\int_{-\infty}^{\infty}\!\frac{d\omega}{2\pi}\!e^{-i\omega(t-t')}\widetilde{\mathcal{G}}(\omega,t_{\text{av}})\!=\!\sum_{m}\!\int_{-\infty}^{\infty}\!\frac{d\omega}{2\pi}\!e^{-i\omega(t-t')}\!e^{-im\Omega\frac{(t+t')}{2}}\widetilde{\mathcal{G}}_{m}(\omega)\!=\!\sum_{m}\!\int_{-\infty}^{\infty}\!\frac{d\omega}{2\pi}\!e^{-i\omega(t-t')} e^{-im\Omega t}\widetilde{\mathcal{G}}_{m}\Big(\omega + \frac{m\Omega}{2}\Big)\!,
\end{equation}
where in the last equality the change of variables $\omega\rightarrow \omega - \frac{m\Omega}{2}$ has been made.
Making use of the property $\int_{-\infty}^{\infty}\frac{d\omega}{2\pi}e^{-i\omega(t-t')}\widetilde{\mathcal{G}}_{m}\Big(\omega + \frac{m\Omega}{2}\Big)= \sum_{n}\int_{0}^{\Omega}\frac{d\omega}{2\pi}e^{-i[\omega+n\Omega](t-t')}\widetilde{\mathcal{G}}_{m}\Big(\omega + \frac{m\Omega}{2} + n\Omega\Big)$, we get
\begin{equation}
\check{\mathcal{G}}(t,t')= \sum_{m,n}\int_{0}^{\Omega}\frac{d\omega}{2\pi}e^{-i[\omega+n\Omega](t-t')}e^{-im\Omega t}\widetilde{\mathcal{G}}_{m}\Big(\omega + \frac{m\Omega}{2} + n\Omega\Big)= \sum_{m,n}\int_{0}^{\Omega}\frac{d\omega}{2\pi}e^{-i(\omega+m\Omega)t}e^{i(\omega + n\Omega)t'}\widetilde{\mathcal{G}}_{m-n}\Big(\omega + \Big(\frac{m+n}{2}\Big)\Omega\Big).
\label{eq3}
\end{equation}
The Floquet representation of the Green's function is then defined in terms of the Wigner oscillating modes as~[\onlinecite{Tsuji2008}]
\begin{equation}
\mathcal{G}_{mn}(\omega) = \widetilde{\mathcal{G}}_{m-n}\Big(\omega + \Big(\frac{m+n}{2}\Big)\Omega\Big).
\label{eq4}
\end{equation}
In the case of the Wigner representation, 
\begin{equation}
\widetilde{\mathcal{G}}_n(\omega) = \int_{-\infty}^{\infty} dt_{\text{rel}} \frac{1}{T}\int_{0}^{T}dt_{\text{av}} \check{\mathcal{G}}(t,t')e^{i\omega t_{\text{rel}} + i n\Omega t_{\text{av}}},
\end{equation}
which depends only on one index $n$, the frequencies aren't restricted. On the other hand, in the Floquet representation the range of $\omega$ is restricted in the ``Brillouin zone", $-\Omega/2 < \omega \leq \Omega/2$, to avoid overcounting. In fact, taking the definition given by Eq. (\ref{eq4})
\begin{equation}
\mathcal{G}_{mn}(\omega) = \int_{-\infty}^{\infty} dt_{\text{rel}} \frac{1}{T}\int_{0}^{T}dt_{\text{av}} \check{\mathcal{G}}(t,t')e^{i(\omega+m\Omega)t} e^{-i(\omega+n\Omega)t'},
\end{equation}
it's easy to check that $\mathcal{G}_{mn}(\omega + l\Omega) = \mathcal{G}_{m+l,n+l}(\omega)$. The advantage of the Floquet representation of the Green's functions relies on the preservation of the multiplication structure in convolution products:
\begin{eqnarray}
\notag
\mathcal{C}(t,t')\!&=&\!\int\!dt''\!\mathcal{A}(t,t'')\mathcal{B}(t'',t')\!=\!\int\!dt''\!\sum_{m,l}\!\sum_{k,n}\!\int_{0}^{\Omega}\!\frac{d\omega}{2\pi}\!\int_{0}^{\Omega}\!\frac{d\omega'}{2\pi}\!e^{-i(\omega+m\Omega)t}e^{i(\omega + l\Omega - \omega' - k\Omega)t''}\mathcal{A}_{ml}(\omega)e^{i(\omega' + n\Omega)t'}\mathcal{B}_{kn}(\omega')\\
\notag
\!&=&\!\sum_{m,n}\sum_{l}\int_{0}^{\Omega}\frac{d\omega}{2\pi} e^{-i(\omega+m\Omega)t}e^{i(\omega + n\Omega)t'}\mathcal{A}_{ml}(\omega)\mathcal{B}_{ln}(\omega)\\
&\!\Leftrightarrow\!&\!\mathcal{C}_{mn}(\omega)\!=\!\sum_{l}\mathcal{A}_{ml}(\omega)\mathcal{B}_{ln}(\omega).
\end{eqnarray}

We are now in position to rewrite the Dyson equation [Eq. (\ref{eq1})], taking into consideration the fact that a function that depends only on the relative time $t-t'$ is block diagonal in the Floquet representation. For example, the unperturbed equilibrium Green's function can be written as
\begin{equation}
\check{g}(t-t')\!=\!\int_{-\infty}^{\infty}\frac{d\omega}{2\pi}e^{-i\omega(t-t')}\widetilde{g}(\omega)=\!\sum_{m}\!\int_{0}^{\Omega}\!\frac{d\omega}{2\pi}e^{-i[\omega+m\Omega](t-t')}\widetilde{g}(\omega + m\Omega)\!=\!\sum_{m,n}\!\int_{0}^{\Omega}\!\frac{d\omega}{2\pi}e^{-i(\omega+m\Omega)t}e^{i(\omega+n\Omega)t'}\widetilde{g}(\omega + m\Omega)\delta_{mn}, 
\end{equation}
meaning that $g_{mn}(\omega) =\widetilde{g}(\omega + m\Omega)\delta_{mn}$. Expressing $V(t'')$ in its Fourier components as $V(t'') = \sum_{l} e^{-il\Omega t''}V_l$, Eq. (\ref{eq1}) in Floquet representation takes the form
\begin{equation}
\mathcal{G}_{mn}(\omega) = \widetilde{g}(\omega + m\Omega)\delta_{mn} + \sum_{l} \widetilde{g}(\omega + m\Omega) V_{m-l}\mathcal{G}_{ln}(\omega), 
\end{equation}
that is to say, a simple matrix product
\begin{equation}
\mathcal{G}(\omega) = g(\omega) + g(\omega) V \mathcal{G}(\omega),
\label{floquetdyson}
\end{equation}
where $V_{ml} = V_{m-l} = \frac{1}{T}\int_{0}^{T}V(t)e^{i(m-l)\Omega t}dt$. In this way, the problem of finding the Green's function has been reduced to an algebraic expression with the matrices defined in an infinite spanned space. Note that if we write the Dyson equation (\ref{eq1}) as
\begin{equation}
[i\partial_t - H(t)]\check{\mathcal{G}}(t,t') = \delta(t-t'),
\end{equation}
with $H(t) = H_0 + V(t)$ and $H_0$ satisfying $(i\partial_t - H_0)\check{g}(t-t') = \delta(t-t')$, then the Green's function in Floquet representation is such that
\begin{equation}
[\omega - H^F]\mathcal{G}(\omega) = \mathcal{I},
\label{floq-dys}
\end{equation}
with $[H_F]_{mn} = (H_0 - m\Omega)\delta_{mn} + V_{m-n}$ the Floquet Hamiltonian.
\subsection{2. Expression for the currents in a periodically driven multi-terminal Josephson junction.}
The Hamiltonian of interest in Nambu representation has the form
\begin{equation}
\hat{H}(t) = \sum_{\bm{k}\nu}\hat{\Psi}^{\dagger}_{\bm{k}\nu}\hat{H}_{\bm{k}\nu}\hat{\Psi}^{}_{\bm{k}\nu} + \hat{\Psi}_{d}^{\dagger}\hat{H}_{d}\hat{\Psi}_d + \sum_{\bm{k}\nu}\Big(\hat{\Psi}^{\dagger}_d\hat{V}_{d\nu}(t)\hat{\Psi}_{\bm{k}\nu}^{} + \hat{\Psi}^{\dagger}_{\bm{k}\nu}\hat{V}_{\nu d}(t)\hat{\Psi}_{d}^{}\Big),
\label{eq11}
\end{equation}
where
\begin{equation}
\hat{H}_{\bm{k}\nu} = \left(\begin{array}{cc}
\varepsilon_{\bm{k}\nu}&-\Delta_{\nu}\\
-\Delta_{\nu}&-\varepsilon_{\bm{k}\nu}
\end{array}\right),\,\,\,\,\,\,\,\,\,\hat{H}_{d} = \left(\begin{array}{cc}
\varepsilon_d&0\\
0&-\varepsilon_{d}
\end{array}\right),\,\,\,\,\,\,\,\,\,\hat{V}_{d\nu}(t) = \left(\begin{array}{cc}
t_{\nu}e^{-i\frac{\varphi_{\nu}(t)}{2}}&0\\
0&-t_{\nu}e^{i\frac{\varphi_{\nu}(t)}{2}}
\end{array}\right) 
\end{equation}
The current flowing into the lead labeled by $\nu$ is expressed as
\begin{equation}
\langle\hat{J}_{\nu}(t)\rangle = -e \frac{i}{\hbar}\langle[\hat{H}(t),\sum_{\bm{k}\sigma}c^{\dagger}_{\bm{k}\nu\sigma}c^{}_{\bm{k}\nu\sigma}]\rangle = i\frac{e}{\hbar}\sum_{\bm{k}\sigma}\Big(t_{\nu}e^{i\frac{\varphi_{\nu}(t)}{2}}\langle c^{\dagger}_{\bm{k}\nu\sigma}(t)d^{}_{\sigma}(t)\rangle-t_{\nu}e^{-i\frac{\varphi_{\nu}(t)}{2}}\langle d^{\dagger}_{\sigma}(t)c^{}_{\bm{k}\nu\sigma}(t)\rangle\Big)
\label{eq27}
\end{equation}
Taking into account the definitions of the propagators
\begin{eqnarray}
\hat{G}^{<}_{d\nu}(\bm{r=0},t,t') &=& i\sum_{\bm{k}}\left(\begin{array}{cc}
\langle c^{\dagger}_{\bm{k}\nu\uparrow}(t') d^{}_{\uparrow}(t)\rangle&\langle c^{}_{\bm{k}\nu\downarrow}(t')d^{}_{\uparrow}(t) \rangle\\
\langle c^{\dagger}_{\bm{k}\nu\uparrow}(t') d^{\dagger}_{\downarrow}(t)\rangle&\langle c^{}_{\bm{k}\nu\downarrow}(t') d^{\dagger}_{\downarrow}(t)\rangle
\end{array}\right)\\
\notag
\hat{G}^{<}_{\nu d}(\bm{r=0},t,t') &=& i\sum_{\bm{k}}\left(\begin{array}{cc}
\langle d^{\dagger}_{\uparrow}(t') c^{}_{\bm{k}\nu\uparrow}(t)\rangle&\langle d^{}_{\downarrow}(t') c^{}_{\bm{k}\nu\uparrow}(t)\rangle\\
\langle d^{\dagger}_{\uparrow}(t') c^{\dagger}_{\bm{k}\nu\downarrow}(t)\rangle&\langle d^{}_{\downarrow}(t') c^{\dagger}_{\bm{k}\nu\downarrow}(t)\rangle
\end{array}\right),
\end{eqnarray}
Eq. (\ref{eq27}) can be written in a more compact form~[\onlinecite{Cuevas1996},\onlinecite{Jauho1996}]
\begin{equation}
\langle\hat{J}_{\nu}(t)\rangle = \frac{e}{\hbar}\text{Tr}\Big\{\sigma_z \Big(\hat{V}_{\nu d}(t)\hat{G}^{<}_{d\nu}(\bm{r=0},t,t)-\hat{V}_{d \nu}(t)\hat{G}^{<}_{\nu d}(\bm{r=0},t,t)\Big)\Big\} = \frac{2e}{\hbar}\Re\text{Tr}\Big\{\sigma_z \Big(\hat{V}_{\nu d}(t)\hat{G}^{<}_{d\nu}(\bm{r=0},t,t)\Big)\Big\},
\label{eq14}
\end{equation}
where in the last equality the fact that $\hat{V}_{d\nu}^{}(t) = \hat{V}_{\nu d}^{\dagger}(t)$ and $\hat{G}^{<}_{\nu d}(\bm{r=0},t,t) = -\hat{G^{<}}^{\dagger}_{d\nu}(\bm{r=0},t,t)$ has been used.
In this problem the time periodicity has been gauged to the superconductor phases $\varphi_{\nu}(t) = \varphi_{\nu}^{0} + b_{\nu}\cos(\Omega t + \chi_{\nu})$. The DC current at each lead is calculated as
\begin{equation}
\langle\hat{J}_{\nu}^{DC}\rangle = \frac{2e}{\hbar}\frac{1}{T}\Re\int_{0}^{T}\text{Tr}\Big\{\sigma_z \Big(\hat{V}_{\nu d}(t)\hat{G}^{<}_{d,\nu}(\bm{r=0},t,t)\Big)\Big\}dt.
\end{equation}
Making use of the Floquet representation of the Green's function it can be written as a trace that only takes into account the zeroth order replica in the infinite matrix representation:
\begin{eqnarray}
\langle\hat{J}_{\nu}^{DC}\rangle &=& \frac{2e}{h}\Re\sum_{n,l}\int_{0}^{\Omega}\text{Tr}\Big\{\sigma_z \Big(\hat{V}_{\nu d, n-l}\widetilde{G}^{<}_{d\nu, l,n}(\bm{r=0},\omega)\Big)\Big\}d\omega\\
\notag
&=& \frac{2e}{h}\Re\sum_{n,l}\int_{n\Omega}^{(n+1)\Omega}\text{Tr}\Big\{\sigma_z \Big(\hat{V}_{\nu d, n-l}\widetilde{G}^{<}_{d\nu, l,n}(\bm{r=0},\omega - n\Omega)\Big)\Big\}d\omega\\
\notag
&=& \frac{2e}{h}\Re\sum_{n,l}\int_{n\Omega}^{(n+1)\Omega}\text{Tr}\Big\{\sigma_z \Big(\hat{V}_{\nu d, n-l}\widetilde{G}^{<}_{d\nu, l-n,0}(\bm{r=0},\omega)\Big)\Big\}d\omega\\
\notag
&=& \frac{2e}{h}\Re\sum_{n,l}\int_{n\Omega}^{(n+1)\Omega}\text{Tr}\Big\{\sigma_z \Big(\hat{V}_{\nu d, 0,l}\widetilde{G}^{<}_{d\nu, l,0}(\bm{r=0},\omega)\Big)\Big\}d\omega=\frac{2e}{h}\Re\sum_{l}\int_{-\infty}^{\infty}\text{Tr}\Big\{\sigma_z \Big(\hat{V}_{\nu d, 0,l}\widetilde{G}^{<}_{d\nu, l,0}(\bm{r=0},\omega)\Big)\Big\}d\omega
\end{eqnarray}
In general, the $m$ oscillating mode of the current, $\langle\hat{J}_{\nu}^{m}\rangle = \frac{1}{T}\int_{0}^{T}e^{im\Omega t}\langle\hat{J}_{\nu}(t)\rangle dt$ can be obtained as
\begin{equation}
\langle\hat{J}_{\nu}^{m}\rangle = \frac{e}{h}\sum_{l}\int_{-\infty}^{\infty}\text{Tr}\Big\{\sigma_z \Big(\hat{V}_{\nu d, m,l}\widetilde{G}^{<}_{d\nu, l,0}(\bm{r=0},\omega) + \hat{V}^{*}_{\nu d, -m,l}\widetilde{G}^{<}{}^{*}_{d\nu, l,0}(\bm{r=0},\omega)\Big)\Big\}d\omega.
\label{eq18}
\end{equation}

In order to calculate $\widetilde{G}^{<}_{d\nu, m,n}(\bm{r=0},\omega)$ we will make use of the Floquet-Dyson equation [Eq. (\ref{floquetdyson})] in matrix form. The unperturbed Green's functions of the dot and the superconducting leads in this representation are given by
\begin{eqnarray}
\widetilde{g}^{r,a}_{dd,mn}(\omega) &=& \delta_{mn}\left(\begin{array}{cc}
\frac{1}{\omega + n\Omega - \varepsilon_d \pm i\eta}&0\\
0&\frac{1}{\omega + n\Omega + \varepsilon_d \pm i\eta}
\end{array}\right)\\
\notag
\widetilde{g}^{r,a}_{\nu\nu,mn}(\bm{r=0},\omega) &=& \frac{\delta_{mn}\rho(\varepsilon_F) \pi}{\sqrt{\Delta_{\nu}^2-(\omega + n\Omega \pm i\eta)^2}}\left(\begin{array}{cc}
-(\omega + n\Omega \pm i\eta)&\Delta_{\nu}\\
\Delta_{\nu}&-(\omega + n\Omega \pm i\eta)
\end{array}\right)
\end{eqnarray}
and the minor functions are simply $\widetilde{g}^{<}_{dd,mn}(\omega) = -\delta_{mn}f(\omega + n\Omega)[\widetilde{g}^{r}_{dd,nn}(\omega)-\widetilde{g}^{a}_{dd,nn}(\omega)]$ and $\widetilde{g}^{<}_{\nu\nu,mn}(\omega) = -\delta_{mn}f(\omega + n\Omega)[\widetilde{g}^{r}_{\nu\nu,nn}(\omega)-\widetilde{g}^{a}_{\nu\nu,nn}(\omega)]$ with $f(\omega)$ the Fermi distribution function.
The matrix elements of the perturbation are
\begin{eqnarray}
\hat{V}_{d\nu,mn} &=& \left(\begin{array}{cc}
t_{\nu}(-i)^{m-n}J_{m-n}(\frac{b_{\nu}}{2})e^{i(-\varphi_{\nu}^{0}/2 + (m-n)\chi_{\nu})}&0\\
0&-t_{\nu}(i)^{m-n}J_{m-n}(\frac{b_{\nu}}{2})e^{i(\varphi_{\nu}^{0}/2 + (m-n)\chi_{\nu})}
\end{array}\right),
\end{eqnarray}
where $J_{m-n}(\frac{b_{\nu}}{2})$ corresponds to the $m-n$-th Bessel function of the first kind. Applying the Langreth's rules, the equation of motion for $\widetilde{G}^{<}_{d\nu}(\bm{r=0},\omega)$ is
\begin{eqnarray}
\widetilde{G}^{<}_{d\nu}(\bm{r=0},\omega) &=& \widetilde{G}^{r}_{dd}(\omega)\hat{V}_{d\nu}(\omega)\widetilde{g}^{<}_{\nu\nu}(\omega) + \widetilde{G}^{<}_{dd}(\omega)\hat{V}_{d\nu}(\omega)\widetilde{g}^{a}_{\nu\nu}(\omega)\\
&=& \widetilde{G}^{r}_{dd}(\omega)\hat{V}_{d\nu}(\omega)\widetilde{g}^{<}_{\nu\nu}(\omega) + \widetilde{G}^{r}_{dd}(\omega)\widetilde{\Sigma}^{<}_{dd}(\omega)\widetilde{G}^{a}_{dd}(\omega)\hat{V}_{d\nu}(\omega)\widetilde{g}^{a}_{\nu\nu}(\omega),
\end{eqnarray}
and the retarded propagator is given by
\begin{equation}
\widetilde{G}^{r}_{dd}(\omega) = [{\widetilde{g}^{r}}{}^{-1}_{dd}(\omega) - \widetilde{\Sigma}^{r}_{dd}(\omega)]^{-1}.
\end{equation}
The dot self-energies are very simple in this case, where the superconductors are treated in BCS mean field theory: 
\begin{equation}
\widetilde{\Sigma}^{<,r,a}_{dd}(\omega) = \sum_{\nu'}\hat{V}_{d\nu'}\widetilde{g}^{<,r,a}_{\nu'\nu'}(\omega)\hat{V}_{\nu' d}.
\end{equation}
\subsection{3. Two time formalism: Quasi-periodically driven Hamiltonians and separation of time scales.}\label{sec2}
When the Hamiltonian of interest is not strictly periodic the $(t,t')$ method, also referred in the literature as the two-time formalism is a useful approach~[\onlinecite{Peskin1993,Pfeifer1983,Drese1999}]. It consists of an extension of the Schr$\ddot{\text{o}}$dinger equation that explictly separates two time scales
\begin{equation}
[i\partial_t + i\partial_{\tau}]\Psi(t,\tau) = H(t,\tau)\Psi(t,\tau).
\end{equation}
The physical state is recovered by equating both times $\Psi(t) = \Psi(t,\tau)\Big\rvert_{t=\tau}$, since
\begin{eqnarray}
i\partial_t \Psi(t) &=& i\partial_t \Psi(t,\tau)\Big\rvert_{t=\tau} +  i\partial_\tau \Psi(t,\tau)\frac{\partial \tau}{\partial t}\Bigg\rvert_{t = \tau}\\
\notag
&=& i\partial_t \Psi(t,\tau)\Big\rvert_{t=\tau} +  [H(t) - i\partial_\tau] \Psi(t,\tau)\Big\rvert_{t = \tau}\\
\notag
&=& H(t)\Psi(t).
\end{eqnarray}
We will take into account that the Hamiltonian preserves its time periodicity in the $t$ variable, while changing slowly in time with $\tau$. Treating the bias as an adiabatic perturbation, the initially occupied Floquet state evolves in time as
\begin{equation}
|\psi^{\alpha}[t,\bm{\varphi}(\tau)]\ra\ra \approx e^{-i\int^{\tau}\varepsilon_{\alpha}(\tau')d\tau'}e^{i\Gamma_{\alpha}(\tau)}\Bigg[|\phi^{F}_{\alpha}[t,\bm{\varphi}(\tau)]\ra\ra -i\hbar\sum_{\beta\neq\alpha}\frac{\la\la\phi^{F}_{\beta}[\bm{\varphi}(\tau)]|\partial_{\tau}\phi^{F}_{\alpha}[\bm{\varphi}(\tau)]\ra\ra}{\varepsilon_{\alpha}[\bm{\varphi}(\tau)]-\varepsilon_{\beta}[\bm{\varphi}(\tau)]}|\phi^{F}_{\beta}[t,\bm{\varphi}(\tau)]\ra\ra\Bigg],
\label{eq41}
\end{equation}
where $\Gamma_{\alpha}(\tau) = i\int^{\tau}\la\la\phi^{F}_{\alpha}(\tau')|\partial_{\tau'}\phi^{F}_{\alpha}(\tau')\ra\ra d\tau'$ is the Berry phase of the state $\alpha$. It is understood that $|\partial_{\tau}\phi^{F}_{\alpha}[\bm{\varphi}(\tau)]\ra\ra = \sum_{\nu}|\partial_{\nu}\phi^{F}_{\alpha}[\bm{\varphi}(\tau)]\ra\ra\dot{\varphi}_{\nu}(\tau)$. This approximation is expected to remain accurate as long as $V/\text{min}|\varepsilon_1(\bm{\varphi})-\varepsilon_2(\bm{\varphi})|\ll1$. Making use of the fact that
\begin{equation}
\la\la\phi^{F}_{\alpha}(\tau)|\hat{J}_{\nu}(\tau)|\phi^{F}_{\beta}(\tau)\ra\ra = \frac{2e}{\hbar}\la\la\phi^{F}_{\alpha}(\tau)|\frac{\partial\hat{H}^{F}(\tau)}{\partial\varphi_{\nu}}|\phi^{F}_{\beta}(\tau)\ra\ra
\end{equation}
and the validity of the Hellman-Feynman theorem for the Floquet states defined within the FFZ,
\begin{equation}
\la\la\phi_{\alpha}^{F}|\frac{\partial\hat{H}_{F}}{\partial \varphi_{\nu}}|\phi_{\beta}^{F}\ra\ra = \frac{\partial \varepsilon_{\alpha}}{\partial\varphi_{\nu}}\delta_{\alpha\beta} + \la\la\phi^{F}_{\alpha}|\partial_{\varphi_{\nu}}\phi^{F}_{\beta}\ra\ra(\varepsilon_{\beta}-\varepsilon_{\alpha}),
\end{equation}
the mean value of the current operator at each lead under the approximation of Eq. (\ref{eq41}) can be expressed as the sum of two contributions:

\begin{equation}
\la\hat{J}_{\nu}[\bm{\varphi}(\tau)]\ra_{\alpha} = \la\la\psi^{\alpha}(\tau)|\hat{J}_{\nu}(\tau)|\psi^{\alpha}(\tau)\ra\ra = \frac{2 e}{\hbar} \frac{\partial\varepsilon_{\alpha}}{\partial \varphi_{\nu}}[\bm{\varphi}(\tau)]-i 2e\sum_{\nu'}\Bigg[\la\la\partial_{\varphi_{\nu}}\phi^{F}_{\alpha}|\partial_{\varphi_{\nu'}}\phi^{F}_{\alpha}\ra\ra - \la\la\partial_{\varphi_{\nu'}}\phi^{F}_{\alpha}|\partial_{\varphi_{\nu}}\phi^{F}_{\alpha}\ra\ra\Bigg]\dot{\varphi}_{\nu'}.
\label{eq10}
\end{equation} 

\subsection{4. Winding number in terms of the Green functions of a non-interacting Hamiltonian.}
The currents flowing into the leads of an N-terminal Josephson junction in the adiabatic regime carry information on the Berry curvature of the ground state of the system. In order to calculate this magnitude taking into account the quasi-particle excitations of the continuum we appeal to the winding number
\begin{equation}
N_2 = \frac{1}{8\pi^2}\int d^2\bm{\varphi} d\omega \text{Tr}\Big[\epsilon^{ij}{\mathcal{G}^t}^{-1}\partial_{\omega}\mathcal{G}^{t}\cdot{\mathcal{G}^{t}}^{-1}\partial_{\varphi_i}\mathcal{G}^{t}\cdot{\mathcal{G}^{t}}^{-1}\partial_{\varphi_j}\mathcal{G}^{t}\Big],
\label{eq45}
\end{equation}
with $\mathcal{G}^{t}$ the time ordered Green's function of the complete Hamiltonian. This correlator can be written in Lehmann representation as
\begin{equation}
\mathcal{G}^{t}(\omega,\bm{\varphi}) = \sum_{n}\frac{\ket{n}\bra{n}}{\omega - \Big(E_{n}(\bm{\varphi}) - \varepsilon_F\Big) + i\eta\text{sgn}\Big(E_{n}(\bm{\varphi}) - \varepsilon_F\Big)}.
\end{equation}
In order to manipulate the expression defined by Eq. (\ref{eq45}) we will make use of the derivative of the inverse of an operator $\partial_{k}A^{-1} = -A^{-1}\cdot\partial_{k}A\cdot A^{-1}$, which in our case traduces to the relations
\begin{eqnarray}
\partial_{\omega}\mathcal{G}^{t} &=& -\mathcal{G}^{t}\cdot\mathcal{G}^{t}\\ 
\notag
\partial_{\varphi_{i}}\mathcal{G}^{t} &=& \mathcal{G}^{t}\cdot\partial_{\varphi_i}\mathcal{H}\cdot\mathcal{G}^{t}.
\end{eqnarray}
The winding number can then be written as
\begin{eqnarray}
N_2 &=& -\frac{1}{8\pi^2}\int d^2\bm{\varphi} d\omega \text{Tr}\Big[\epsilon^{ij}\mathcal{G}^{t}\cdot\partial_{\varphi_i}\mathcal{H}\cdot\mathcal{G}^{t}\cdot\partial_{\varphi_j}\mathcal{H}\cdot\mathcal{G}^{t}\Big]\\
\notag
&=&-\frac{1}{8\pi^2}\sum_{m,n}\int d^2\bm{\varphi} d\omega\epsilon^{ij}\frac{\bra{n}\partial_{\varphi_i}\mathcal{H}\ket{m}\bra{m}\partial_{\varphi_j}\mathcal{H}\ket{n}}{\Big[\omega - \Big(E_{n}(\bm{\varphi}) - \varepsilon_F\Big) + i\eta\text{sgn}\Big(E_{n}(\bm{\varphi}) - \varepsilon_F\Big)\Big]^2\Big[\omega - \Big(E_{m}(\bm{\varphi}) - \varepsilon_F\Big) + i\eta\text{sgn}\Big(E_{m}(\bm{\varphi}) - \varepsilon_F\Big)\Big]}.
\end{eqnarray}
The integral over frequencies can be done by looking at the poles of the ingrand. It has a simple pole and one of second order with residues:
\begin{eqnarray}
\text{Res}[\omega_p^{(1)} = E_m - \varepsilon_F - i\eta\text{sgn}\Big(E_{m}(\bm{\varphi}) - \varepsilon_F\Big)] &=& -\frac{1}{8\pi^2}\sum_{m,n}\int d^2\bm{\varphi}\epsilon^{ij}\frac{\bra{n}\partial_{\varphi_i}\mathcal{H}\ket{m}\bra{m}\partial_{\varphi_j}\mathcal{H}\ket{n}}{[E_m - E_n]^2}\\
\notag
\text{Res}[\omega_p^{(2)} = E_n - \varepsilon_F - i\eta\text{sgn}\Big(E_{n}(\bm{\varphi}) - \varepsilon_F\Big)] &=& \frac{1}{8\pi^2}\sum_{m,n}\int d^2\bm{\varphi}\epsilon^{ij}\frac{\bra{n}\partial_{\varphi_i}\mathcal{H}\ket{m}\bra{m}\partial_{\varphi_j}\mathcal{H}\ket{n}}{[E_m - E_n]^2}
\end{eqnarray}
The integral is finite if and only if $\text{sgn}\Big(E_{n}(\bm{\varphi}) - \varepsilon_F\Big)\neq \text{sgn}\Big(E_{m}(\bm{\varphi}) - \varepsilon_F\Big)$. Suming up the contributions where $E_n < \varepsilon_F$ \& $E_m > \varepsilon_F$ and viceversa we finally obtain
\begin{equation}
N_2 = \frac{i}{2\pi}\sum\limits_{\substack{m\epsilon\text{desocc}\\n\epsilon\text{occ}}}\int d^2\bm{\varphi}\epsilon^{ij}\frac{\bra{n}\partial_{\varphi_i}\mathcal{H}\ket{m}\bra{m}\partial_{\varphi_j}\mathcal{H}\ket{n}}{[E_m - E_n]^2} = \frac{1}{2\pi}\sum\limits_{n\epsilon \text{occ}}\int d^2\bm{\varphi}\mathcal{F}_{ij}^{n}(\bm{\varphi}) = \sum\limits_{n\epsilon \text{occ}} \mathcal{C}_n,
\end{equation}
which means that the kernel of the invariant, the Berry curvature of the ground state, can be generally writen as
\begin{equation}
\mathcal{F}_{ij}^{g} = \frac{1}{4\pi}\int_{-\infty}^{\infty}d\omega\text{Tr}\Big[\epsilon^{ij}{\mathcal{G}^{(0)}}^{-1}\partial_{\varphi_{i}}\mathcal{G}^{(0)}\cdot{\mathcal{G}^{(0)}}^{-1}\partial_{\varphi_{j}}\mathcal{G}^{(0)}\cdot{\mathcal{G}^{(0)}}^{-1}\partial_{\omega}\mathcal{G}^{(0)}\Big].
\label{eq46}
\end{equation}
In our particular case, this ``generalized" Berry curvature can be written in terms of the Green's functions calculated at the link of the dot and the superconducting leads. In fact, the quantity of interest is found to be
\begin{eqnarray}
\notag
\!\sum\limits_{n\epsilon \text{occ}}\mathcal{F}_{\mu\rho}^{n}(\bm{\varphi})\!&=&\!\frac{1}{4\pi}\int d\omega \epsilon^{\mu\rho}\text{Tr}\Big[\partial_{\omega}\mathcal{G}^{t}_{d\mu}\frac{\partial V_{\mu d}}{\partial \varphi_{\mu}}\mathcal{G}^{t}_{d\rho}\frac{\partial V_{\rho d}}{\partial \varphi_{\rho}} + \partial_{\omega}\mathcal{G}^{t}_{\rho\mu}\frac{\partial V_{\mu d}}{\partial \varphi_{\mu}}\mathcal{G}^{t}_{dd}\frac{\partial V_{d\rho}}{\partial \varphi_{\rho}}+\partial_{\omega}\mathcal{G}^{t}_{dd}\frac{\partial V_{d\mu }}{\partial \varphi_{\mu}}\mathcal{G}^{t}_{\mu\rho}\frac{\partial V_{\rho d}}{\partial \varphi_{\rho}}+\partial_{\omega}\mathcal{G}^{t}_{\rho d}\frac{\partial V_{d\mu }}{\partial \varphi_{\mu}}\mathcal{G}^{t}_{\mu d}\frac{\partial V_{d\rho}}{\partial \varphi_{\rho}}\Big],\\
\label{eq36}
\end{eqnarray}
where the trace is in Nambu space.
\subsection{5. Wigner representation and adiabatic expansion of the Green' s functions.}
In general, the adiabatic expansion can be done directly on the Green's function of the Hamiltonian. We'll begin with the equation of motion of the Green's functions
\begin{equation}
[i\partial_t - H(t)]\mathcal{G}(t,t') = \delta(t-t') + \int dt_1 \Sigma(t,t_1)\mathcal{G}(t_1,t'),
\label{eq53}
\end{equation}
where $H(t)$ is a one-particle time dependent ``solvable" hamiltonian and $\Sigma(t,t_1)$ is the self-energy. The approach is based on the separation of slow and fast timescales in the Dyson equation by taking advantage of the Wigner representation. Under this formalism, two time variables, the relative time $t_{\text{rel}}=t-t'$ and the average one $t_{\text{av}}=\frac{t+t'}{2}$ are handled. Time derivatives of the former are used as a small parameter, making feasible a perturbation scheme. The Wigner representation is given by
\begin{eqnarray}
\notag
\mathcal{G}(t,t') &=& \int_{-\infty}^{\infty}\frac{d\omega}{2\pi}e^{-i\omega (t-t')}\widetilde{\mathcal{G}}(\omega,t_{\text{av}})\\
\widetilde{\mathcal{G}}(\omega,t_{\text{av}}) &=& \int_{-\infty}^{\infty}dt_{\text{rel}}e^{i\omega t_{\text{rel}}}\mathcal{G}(t,t').
\end{eqnarray}
In what's next we will take into account that $\partial_t = \frac{1}{2}\partial_{t_{\text{av}}} + \partial_{t_{\text{rel}}}$ and make use of the Moyal product $\star$ for the convolution of two functions 
\begin{equation}
\int_{-\infty}^{\infty} e^{i\omega t_{\text{rel}}}A(t,t_1)B(t_1,t')dt_{\text{rel}} = e^{\frac{1}{2i}[\partial_{t_{\text{av}}}^A\partial_{\omega}^{B}-\partial_{\omega}^{A}\partial_{t_{\text{av}}}^B]}\widetilde{A}(\omega,t_{\text{av}})\widetilde{B}(\omega,t_{\text{av}}) = \widetilde{A}(\omega,t_{\text{av}})\star\widetilde{B}(\omega,t_{\text{av}})
\end{equation}
Eq. (\ref{eq53}) can be written to all orders as~[\onlinecite{Kershaw2017}]
\begin{equation}
[\omega + \frac{i}{2}\partial_{\tav} - e^{\frac{1}{2i}\partial_{\omega}^{\mathcal{G}}\partial_{\tav}^{H}}H(\tav)]\widetilde{\mathcal{G}}(\omega,\tav) = \mathcal{I} + e^{\frac{1}{2i}[\partial_{t_{\text{av}}}^{\Sigma}\partial_{\omega}^{\mathcal{G}}-\partial_{\omega}^{\Sigma}\partial_{t_{\text{av}}}^{\mathcal{G}}]}\widetilde{\Sigma}(\omega,t_{\text{av}})\widetilde{\mathcal{G}}(\omega,t_{\text{av}}),
\label{eq56}
\end{equation} 
where we applied the following manipulation
\begin{eqnarray}
\notag
\int_{-\infty}^{\infty}e^{i\omega\trel}H(\tav + \trel/2)\mathcal{G}\big(\tav+\frac{\trel}{2}, \tav - \frac{\trel}{2}\big)d\trel &=& \int_{-\infty}^{\infty}e^{i\omega\trel}e^{\frac{\trel}{2}\partial_{\tav}^{H}}H(\tav)\mathcal{G}\big(\tav+\frac{\trel}{2}, \tav - \frac{\trel}{2}\big)d\trel\\
\notag
&=& \int_{-\infty}^{\infty}e^{\frac{1}{2 i}\partial_{\omega}\partial_{\tav}^{H}}e^{i\omega\trel} H(\tav)\mathcal{G}\big(\tav+\frac{\trel}{2}, \tav - \frac{\trel}{2}\big)d\trel\\
&=& e^{\frac{1}{2i}\partial_{\omega}^{\mathcal{G}}\partial_{\tav}^{H}}H(\tav)\widetilde{\mathcal{G}}(\omega,\tav).
\end{eqnarray}

We are interested in a first order approximation, where both the Green's functions and the selfenergy are perturbed up to first derivatives of the average time. In that case, Eq. (\ref{eq56}) is given by
\begin{equation}
\Big[\omega + \frac{i}{2}\partial_{\tav} - H(\tav) - \frac{1}{2i}\partial_{\tav}H(\tav)\partial_{\omega}\Big]\widetilde{\mathcal{G}}(\omega,\tav)\!=\!\mathcal{I} + \widetilde{\Sigma}(\omega,t_{\text{av}})\widetilde{\mathcal{G}}(\omega,t_{\text{av}})\!+\!\frac{1}{2i}\Big(\partial_{\tav}\widetilde{\Sigma}(\omega,t_{\text{av}})\partial_{\omega}\widetilde{\mathcal{G}}(\omega,t_{\text{av}})\!-\! \partial_{\omega}\widetilde{\Sigma}(\omega,t_{\text{av}})\partial_{\tav}\widetilde{\mathcal{G}}(\omega,t_{\text{av}})\Big) 
\end{equation}
Introducing a series expansion for the Green function as well as for the selfenergy $\widetilde{\mathcal{G}}(\omega,t_{\text{av}}) = \sum_n\widetilde{\mathcal{G}}^{(n)}$ and $\widetilde{\Sigma}(\omega,t_{\text{av}}) = \sum_n\widetilde{\Sigma}^{(n)}$ we can obtain the system of equations
\begin{eqnarray}
\label{eq6}
\Big[\omega - H - \widetilde{\Sigma}^{(0)}\Big]\widetilde{\mathcal{G}}^{(0)} &=& \mathcal{I}\\
\label{eq7}
\Big[\omega - H - \widetilde{\Sigma}^{(0)}\Big]\widetilde{\mathcal{G}}^{(1)} &=& \widetilde{\Sigma}^{(1)}\widetilde{\mathcal{G}}^{(0)} + \frac{1}{2i}\partial_{\tav}H\partial_{\omega}\widetilde{\mathcal{G}}^{(0)} + \frac{1}{2i}\partial_{\tav}\widetilde{\mathcal{G}}^{(0)} + \frac{1}{2i}\Big(\partial_{\tav}\widetilde{\Sigma}^{(0)}\partial_{\omega}\widetilde{\mathcal{G}}^{(0)} - \partial_{\omega}\widetilde{\Sigma}^{(0)}\partial_{\tav}\widetilde{\mathcal{G}}^{(0)}\Big),
\end{eqnarray}
where the arguements $(\omega,\tav)$ are omitted for the sake of brevity. Solving Eq. (\ref{eq6}) and replacing into Eq. (\ref{eq7}) we get
\begin{eqnarray}
\label{eq8}
\widetilde{\mathcal{G}}^{(0)} &=& \Big[\omega - H - \widetilde{\Sigma}^{(0)}\Big]^{-1}\\
\label{eq9}
\widetilde{\mathcal{G}}^{(1)} &=& \widetilde{\mathcal{G}}^{(0)}\widetilde{\Sigma}^{(1)}\widetilde{\mathcal{G}}^{(0)} + \frac{1}{2i}\widetilde{\mathcal{G}}^{(0)}\Bigg[(\mathcal{I} - \partial_{\omega}\widetilde{\Sigma}^{(0)})\widetilde{\mathcal{G}}^{(0)}, (\partial_{\tav}\widetilde{\Sigma}^{(0)} + \partial_{\tav}H)\widetilde{\mathcal{G}}^{(0)}\Bigg],
\end{eqnarray}
where $\Big[.,.\Big]$ is the commutator. We've also used the properties of the derivatives of inverse functions $\partial_{\lambda}A^{-1} = -A^{-1}\partial_{\lambda}A A^{-1}$, meaning in this case that
\begin{eqnarray}
\partial_\omega\widetilde{\mathcal{G}}^{(0)} &=& -\widetilde{\mathcal{G}}^{(0)}[\mathcal{I} - \partial_{\omega}\widetilde{\Sigma}^{(0)}]\widetilde{\mathcal{G}}^{(0)}\\
\partial_{\tav}\widetilde{\mathcal{G}}^{(0)} &=& \widetilde{\mathcal{G}}^{(0)}(\partial_{\tav}\widetilde{\Sigma}^{(0)} + \partial_{\tav}H)\widetilde{\mathcal{G}}^{(0)}
\end{eqnarray}

In BCS approximation the Hamiltonian is quadratic, so the junction can be treated with a one particle scheme, meaning the interacting selfenergy is zero. The mean value of a single particle operator $\hat{\mathcal{O}}(t) = \sum_{\alpha\beta}\hat{\Psi}^{\dagger}_{\beta}(t)\mathcal{O}_{\beta\alpha}\hat{\Psi}_{\alpha}(t)$ is given by
\begin{eqnarray}
\langle\phi_{0}|\hat{\mathcal{O}}(t)|\phi_{0}\rangle = \lim_{t'-t\to\epsilon^{+}}\sum_{\alpha\beta}\mathcal{O}_{\beta\alpha}\langle\phi_{0}|\hat{\Psi}^{\dagger}_{\beta}(t')\hat{\Psi}_{\alpha}(t)|\phi_{0}\rangle = -i\lim_{t'-t\to\epsilon^{+}}\sum_{\alpha\beta}\mathcal{O}_{\beta\alpha}\mathcal{G}_{\alpha\beta}(t,t') = -i\lim_{t'-t\to\epsilon^{+}}\text{Tr}\Big[\mathcal{O}\mathcal{G}(t,t')\Big]
\label{eq13}
\end{eqnarray}
By writing Eq. (\ref{eq13}) in Wigner represention and expanding the Green's function up to first order we obtain
\begin{equation}
\langle\mathcal{O}(t)\rangle \simeq -i\lim_{t'-t\to\epsilon^{+}}\int_{-\infty}^{\infty}\frac{d\omega}{2\pi} \text{Tr}\Big[\mathcal{O}\mathcal{G}^{(0)}\Big] + \lim_{t'-t\to\epsilon^{+}}\int_{-\infty}^{\infty}\frac{d\omega}{4\pi}\text{Tr}\Big[\partial_{\tav}H\mathcal{G}^{(0)}\mathcal{O}\partial_{\omega}\mathcal{G}^{(0)} - \mathcal{O}\mathcal{G}^{(0)}\partial_{\tav}H\partial_{\omega}\mathcal{G}^{(0)}\Big].
\end{equation}
By specifying $\mathcal{O}$ as the current operator $\mathcal{J}_{\mu} = 2e\partial_{\varphi_{\mu}}H$ and using that $\lim_{t'-t\to\epsilon^{+}}\partial_{\tav}H = \sum_{\nu}\partial_{\varphi_{\nu}}H\dot{\varphi}_{\nu}(t)$, we get that the first order correction to the current is given by
\begin{eqnarray}
\langle\mathcal{J}^{(1)}_{\mu}(t)\rangle &=& 2e\sum_{\nu}\int_{-\infty}^{\infty}\frac{d\omega}{4\pi}\text{Tr}\Big[\epsilon^{\nu\mu}\partial_{\varphi_{\nu}}H\mathcal{G}^{(0)}\partial_{\varphi_{\mu}}H\partial_{\omega}\mathcal{G}^{(0)}\Big]\dot{\varphi}_{\nu}(t)\\
&=& -2e\sum_{\nu}\int_{-\infty}^{\infty}\frac{d\omega}{4\pi}\text{Tr}\Big[\epsilon^{\mu\nu}{\mathcal{G}^{(0)}}^{-1}\partial_{\varphi_{\mu}}\mathcal{G}^{(0)}\cdot{\mathcal{G}^{(0)}}^{-1}\partial_{\varphi_{\nu}}\mathcal{G}^{(0)}\cdot{\mathcal{G}^{(0)}}^{-1}\partial_{\omega}\mathcal{G}^{(0)}\Big]\dot{\varphi}_{\nu}(t)\\
&=&-2e\sum_{\nu}\mathcal{F}_{\mu\nu}\dot{\varphi}_{\nu}(t),
\end{eqnarray}
where in the last equality we used the invariant defined in Eq. (\ref{eq46}).

\end{widetext}

%

\end{document}